\documentclass[english,12pt]{article}
\usepackage{array}
\usepackage{graphicx}
\usepackage{amssymb}
\usepackage{amsmath}
\usepackage{multirow}
\usepackage{prettyref}
\usepackage{babel}
\usepackage{units}
\usepackage[latin1]{inputenc}
\usepackage{amsfonts}
\usepackage{amssymb}
\usepackage{babel}
\usepackage{cite}
\def\@fmsl@sh#1#2#3{\m@th\ooalign{$\hfil#1\mkern#2/\hfil$\crcr$#1#3$}}
 \def\eq#1\en{\begin{equation}#1\end{equation}}
\def\s[#1,#2]{[#1\stackrel{\star}{,}#2]}
\def\sx[#1,#2]{[#1\stackrel{\star_{x}}{,}#2]}

\newcommand{\nc}{\newcommand}
\nc{\beq}{\begin{equation}}
\nc{\eeq}{\end{equation}}
\nc{\beqa}{\begin{eqnarray}}
\nc{\eeqa}{\end{eqnarray}}

\def\bc{\begin{center}}
\def\ec{\end{center}}

\def\to{\rightarrow}

\def\gsim{\mathrel{\mathpalette\atversim>}}

\def\bc{\begin{center}}
\def\ec{\end{center}}

\def\gsim{\mathrel{\rlap{\lower4pt\hbox{\hskip1pt$\sim$}}

    \raise1pt\hbox{$>$}}}       %greater than or approx. symbol

\def\gsim{\mathrel{\rlap{\lower4pt\hbox{\hskip1pt$\sim$}}
    \raise1pt\hbox{$>$}}}       %greater than or approx. symbol

\begin{document}
\makeatletter
\def\fmslash{\@ifnextchar[{\fmsl@sh}{\fmsl@sh[0mu]}}
\def\fmsl@sh[#1]#2{%
  \mathchoice
    {\@fmsl@sh\displaystyle{#1}{#2}}%
    {\@fmsl@sh\textstyle{#1}{#2}}%
    {\@fmsl@sh\scriptstyle{#1}{#2}}%
    {\@fmsl@sh\scriptscriptstyle{#1}{#2}}}
\def\@fmsl@sh#1#2#3{\m@th\ooalign{$\hfil#1\mkern#2/\hfil$\crcr$#1#3$}}
\makeatother
%\baselineskip 24pt

%%%%%%%%%%%%%%%%%%%%%%%%%%%%%%%%%%%%%%%%%%%%%%%%%%%%%%%%%%%%%%%%%
%%%
%%%                      TITLE PAGE
%%%
%%%%%%%%%%%%%%%%%%%%%%%%%%%%%%%%%%%%%%%%%%%%%%%%%%%%%%%%%%%%%%%%%
\thispagestyle{empty}
\begin{titlepage}
\boldmath
\begin{center}
  \Large {\bf Quantum Corrections to the Gravitational Backreaction}
    \end{center}
\unboldmath
\vspace{0.2cm}
\begin{center}
{{\large Iber\^e Kuntz}\footnote{ibere.kuntz@sussex.ac.uk}}
 \end{center}
\begin{center}
{\sl Physics $\&$ Astronomy, 
University of Sussex,   Falmer, Brighton, BN1 9QH, United Kingdom 
}
\end{center}
\vspace{5cm}
\begin{abstract}
\noindent Effective Field Theory techniques are used to study the leading order quantum corrections to the gravitational wave backreaction. The effective stress-energy tensor is calculated and it is shown that it has a non-vanishing trace that contributes to the cosmological constant. By comparing the result obtained with LIGO's data, the first bound on the amplitude of the massive mode is found: $\epsilon < 1.4\times 10^{-33}$.
\end{abstract}  
\end{titlepage}

%\pacs{}

%%%%%%%%%%%%%%%%%%%%%%%%%%%%%%%%%%%%%%%%%%%%%%%%%%%%%%%%%%%%%%%%
%%%
%%%                     INTRODUCTION
%%%
%%%%%%%%%%%%%%%%%%%%%%%%%%%%%%%%%%%%%%%%%%%%%%%%%%%%%%%%%%%%%%%%

\newpage
\section{Introduction}
The recent experimental discovery of gravitational waves (GWs) \cite{Abbott:2016blz} has marked a new era for both observational and theoretical physics. With the new coming data from LIGO and from future experiments like LISA, it will become possible to test modified gravity theories, establishing for which range of parameters these theories agree with observations. Particularly, it may be even possible to test Quantum Gravity in its low energy limit, even though a complete quantum theory for gravity remains one of the greatest problems in modern physics.

A natural observable to consider is the GW energy. As a non-linear phenomenon, gravity couples to itself and thus gravitates, which means that GWs --- being a manifestation of gravity --- produce a backreaction into the spacetime. Hence, one should be able to find a stress-energy tensor for the GWs that accounts for this phenomenon. In the case of classical General Relativity (GR), such a stress-energy tensor is known:
\begin{equation}
t_{\mu\nu}^{\text{GR}} = \frac{1}{32\pi G}\left<\partial_\mu h_{\alpha\beta}\partial_\nu h^{\alpha\beta}\right>,
\label{eq:gr}
\end{equation}
where the brackets denote an average over spacetime, which is responsible for taking only the long-wavelength modes; its precise definition will be explained later on. The GW stress-energy tensor has also been calculated for some other theories, including $f(R)$, Chern-Simons and higher-derivative gravity \cite{Stein:2010pn,Preston:2016sip,Preston:2014tua,Saito:2012xa}. In \cite{Berry:2011pb},  it was indicated how the parameters of an analytic $f(R)$ theory could be constrained by the measurement of the energy or momentum carried away by the GWs.

The phenomenology, however, is not the only motivation. An alternative for dark energy has been proposed based on the effective stress-energy tensor \cite{Preston:2014tua,Rasanen:2010fe,Rasanen:2003fy,Buchert:2011sx}. Although this is not possible in GR because of the vanishing trace of $t_{\mu\nu}^{\text{GR}}$, it was pointed out it could be possible in modified gravity theories. However, it was also found that in some models such as Starobinsky gravity, the effective stress-energy tensor could not be the only factor as it does not produce the right value for the cosmological constant \cite{Preston:2014tua}. We will show that the large contributions from the Standard Model cannot be canceled by the quantum gravitational effects, thus requiring the existence of another mechanism able to reconcile the discrepancy between theory and observation.

The purpose of this paper is, then, two-fold: we will establish new phenomenological bounds and discuss the possibility of generating a contribution to the cosmological constant in this framework. Effective Field Theory techniques will be used to calculate quantum contributions to the GW backreaction and to the wave equation in an arbitrary background. The short-wave formalism will be employed, consisting of an averaging procedure that separates the low frequency modes from the high ones, in order to calculate the GW stress-energy tensor in quantum GR. These theoretical findings will be useful to constrain some of the parameters of Effective Quantum Gravity by the direct comparison with LIGO's observations. Furthermore, on the theoretical side, they give us new insights of gravity at the quantum level since this approach is model independent and, as such, leads to genuine predictions of Quantum Gravity.

This paper is organized as follows. In Section \ref{sec:eft}, we will review the main results of the Effective Field Theory approach applied to gravity. In Section \ref{sec:bg}, we use the short-wave formalism to calculate the leading order quantum corrections to the GW stress-energy tensor. The result allows us to constraint the amplitude of the massive mode present in Effective Quantum Gravity. In Section \ref{sec:prop}, we discuss the quantum corrections to the propagation of GWs and we show that the equation describing the propagation in curved spacetime can be obtained by performing a minimal coupling prescription to the equation in Minkowski space. We draw the conclusions in Section \ref{sec:conc}.
\section{Effective Quantum Gravity}
\label{sec:eft}
The quantum effective action of gravity up to quadratic order in curvature is given by \cite{Donoghue:2014yha}
\begin{align}
\Gamma = \int\mathrm{d}^4x\sqrt{-g}\bigg(&\frac{M_p^2}{2}R + b_1 R^2 + b_2 R_{\mu\nu}R^{\mu\nu} + c_1 R\log\frac{\Box}{\mu^2}R + c_2 R_{\mu\nu}\log\frac{\Box}{\mu^2}R^{\mu\nu}\nonumber\\
& + c_3 R_{\mu\nu\rho\sigma}\log\frac{\Box}{\mu^2}R^{\mu\nu\rho\sigma}\bigg),
\label{eq:qaction}
\end{align}
where $M_p = (8\pi G)^{-1/2}$ is the reduced Planck mass, $G$ is the Newton's constant, $\mu$ is the renormalization scale and the kernel $R$ denotes the Riemann tensor and its contractions (Ricci tensor and Ricci scalar) depending on the number of indices it carries. The signature $(-+++)$ will be adopted. We set the bare cosmological constant to zero as it is not important to our considerations. The coefficients $b_i$ are free parameters and must be fixed by observations, while the coefficients $c_i$ are predictions of the infra-red theory and depend on the field content under consideration (see Table 1 in \cite{Donoghue:2014yha} for their precise values). The log operators are known to lead to acausal effects that need to be removed by resolving the non-local operator as
\begin{equation}
\log\frac{\Box}{\mu^2} = \int_0^\infty\mathrm{d}s\left(\frac{1}{\mu^2+s} - G(x,x',\sqrt{s})\right),
\label{eq:log}
\end{equation}
where $G(x,x';\sqrt{s})$ is a Green's function for
\begin{equation}
(\Box + k^2)G(x,x';k) = \delta^4(x-x'),
\end{equation}
and imposing proper boundary conditions on $G(x,x';k)$ so that the result respects causality. Moreover, in the weak field limit, the log terms are not indepedent due to the following relation (see \cite{Preston:2016sip}):
\begin{equation}
\delta\int\mathrm{d}^4x\sqrt{-g}\left(R_{\mu\nu\rho\sigma}\log\frac{\Box}{\mu^2}R^{\mu\nu\rho\sigma}-4 R_{\mu\nu}\log\frac{\Box}{\mu^2}R^{\mu\nu} + R\log\frac{\Box}{\mu^2}R\right) \stackrel{\text{weak}}{=} 0.
\end{equation}
This can also be seen by linearizing the field equations \cite{Calmet:2017rxl}. The log operators in the above expression certainly break the topological invariance given by the Gauss-Bonnet theorem. Nonetheless, such expression still provides a useful relation that can be used to simplify calculations in the weak field limit. Therefore, since we will be interested only in the weak field scenario, the last term in \eqref{eq:qaction} will be eliminated in favour of the other two log terms, which translates into a shift of their coefficients:
\begin{align}
c_1 &\to\alpha \equiv c_1-c_3,\\
c_2 &\to \beta\equiv c_2+4 c_3.
\end{align}
Hence, from now on, $\alpha$ will denote the coefficient of $R\log\frac{\Box}{\mu^2}R$ and $\beta$ the coefficient of $R_{\mu\nu}\log\frac{\Box}{\mu^2}R^{\mu\nu}$. Note, however, that the last term in \eqref{eq:qaction} will give independent contrubutions in the non-linear regime and, in particular, the background equations of motion (left-hand side of \eqref{eq:bkeom} below) will be changed, but none of this affects the right-hand side of \eqref{eq:bkeom}.

The quantum action \eqref{eq:qaction} yields the equations of motion (EOM)
\begin{equation}
G_{\mu\nu} + \Delta G_{\mu\nu}^L + \Delta G_{\mu\nu}^{NL} = 8\pi G T_{\mu\nu},
\label{eq:eom}
\end{equation}
where $\Delta G_{\mu\nu}^L$ denotes the local contribution to the modification of Einstein's tensor and $\Delta G_{\mu\nu}^{NL} = \Delta G_{\mu\nu}^\alpha + \Delta G_{\mu\nu}^\beta$ is the non-local one (due to the log operator), coming from the terms proportional to $\alpha$ and $\beta$, denoted by $\Delta G_{\mu\nu}^\alpha$ and $\Delta G_{\mu\nu}^\beta$, respectively. Here we will show only the calculation of the non-local part $\Delta G_{\mu\nu}^{NL}$ as the local contribution can be straightforwadly obtained from it. However, our final results will be completely general, including both the local and non-local physics. The $\Delta G_{\mu\nu}^\alpha$ has been calculated in the literature \cite{Codello:2015pga}:
\begin{equation}
\label{eq:mod}
-\xi\Delta G_{\mu\nu}^\alpha = 2\left(R_{\mu\nu}-\frac{1}{4}g_{\mu\nu}R\right)\!\!\left(\log\frac{\square}{\mu^{2}}R\right)-2 \left(\nabla_{\mu}\nabla_{\nu}-g_{\mu\nu}\square\right) \left(\log\frac{\square}{\mu^{2}}R\right),
\end{equation}
where $\xi = \frac{1}{16\pi G\alpha}$. Note that the integral term appearing in \cite{Codello:2015pga}, which comes from the variation of the D'Alembert operator, is not present here. This is because in the weak field limit the variation of the D'Alembert operator leads to negligible contributions \cite{Donoghue:2015nba}. The other contribution to $\Delta G_{\mu\nu}$ is given by
\begin{align}
\label{eq:mod1}
\zeta\Delta G_{\mu\nu}^\beta &= -\frac12 g_{\mu\nu}R_{\rho\sigma}\log\left(\frac{\Box}{\mu^2}\right)R^{\rho\sigma} + \Box\log\left(\frac{\Box}{\mu^2}\right)R_{\mu\nu}\nonumber + g_{\mu\nu}\nabla_\rho\nabla_\sigma\log\left(\frac{\Box}{\mu^2}\right)R^{\rho\sigma}\\
& + R_\mu^\sigma\log\left(\frac{\Box}{\mu^2}\right)R_{\nu\sigma} + R_\nu^\sigma\log\left(\frac{\Box}{\mu^2}\right)R_{\mu\sigma}\\
& - \nabla_\rho\nabla_\mu \log\left(\frac{\Box}{\mu^2}\right)R^{\rho}_{\nu} - \nabla_\rho\nabla_\nu \log\left(\frac{\Box}{\mu^2}\right)R_{\mu}^{\rho}\nonumber
\end{align}
where $\zeta = \frac{1}{16\pi G\beta}$.

\section{Gravitational wave backreaction}
\label{sec:bg}
The first step is to separate the fluctuations $h_{\mu\nu}$ (GWs) from the background geometry $\bar g_{\mu\nu}$, via $g_{\mu\nu} = \bar g_{\mu\nu} + h_{\mu\nu}$. This separation is only meaningful in the limit where the GW wavelength $\lambda$ is much smaller than the background radius $L$, i.e. $\lambda\ll L$, so that a clear distinction between background and GW can be made. As a first approximation, the background metric $\bar g_{\mu\nu}$ will be used to raise/lower indices as well as to build all the operators, e.g. $\Box = \bar g^{\mu\nu}\nabla_\mu\nabla_\nu$. The connection is also assumed to be compatible with $\bar g_{\mu\nu}$ instead of $g_{\mu\nu}$.

The separation of gravity into background and fluctuations allows one to expand the Ricci tensor as
\begin{equation}
R_{\mu\nu} = \bar R_{\mu\nu} + R_{\mu\nu}^{(1)} + R_{\mu\nu}^{(2)} + O(h^3),
\end{equation}
where the bar quantities are calculated with respect to the background and the rest depends only on the fluctuation. The superscript $(n)$ is used to indicate the order in $h$ of the underlying term. Naively, one could think that the EOM could be calculated order by order in $h$, giving no backreaction into the background. The problem is that there are two small parameters in the game, namely the fluctuation amplitude $h$ and $\varepsilon\equiv \frac{\lambda}{L}$, so that one can compensate the other. Their relation is fixed by the EOM\footnote{Note that $\bar R_{\mu\nu}\sim \frac{1}{L^2}$, $R_{\mu\nu}^{(n)}\sim \frac{h^{n}}{\lambda^2}$ and the contribution of GWs to the curvature is negligible compared to the contribution of matter sources.} and in the presence of external matter
\begin{equation}
h\ll\varepsilon\ll 1,
\label{eq:con}
\end{equation}
as can be seen from Equation \eqref{eq:eom}.

To obtain the GW backreaction, one then needs to calculate the average of tensor fields over a region of length scale $d$, where $\lambda\ll d\ll L$. This makes the high-frequency modes go away due to their rapid oscillation, but leave the low modes intact. The subtle point is that there is no canonical way of summing tensors based on different points of a manifold. Here the Isaacson's definition \cite{Isaacson:1968zza,Isaacson:1967zz} of the average of a tensor will be used, which is based on the idea of parallel transporting the tensor field along geodesics from each spacetime position to a common point where its different values can be compared:
\begin{equation}
\left<A_{\mu\nu}(x)\right> = \int j_\mu^{\alpha'}(x,x') j_\nu^{\beta'}(x,x') A_{\alpha'\beta'}(x')f(x,x')\sqrt{-\bar g(x')}d^4x',
\end{equation}
where $j_\mu^{\alpha'}$ is the bivector of geodesic parallel displacement and $f(x,x')$ is a weight function that falls quickly and smoothly to zero when $|x-x'|>d$, such that
\begin{equation}
\int_\text{all space} f(x,x')\sqrt{-\bar g(x')}d^4x'=1.
\end{equation}
From this definition, the following rules can be proven \cite{Stein:2010pn}:
\begin{itemize}
\item The average of an odd product of short-wavelength quantities vanishes.
\item The derivative of a short-wavelength tensor averages to zero, e.g., $\left<\nabla_\mu T_{\alpha\beta}^\mu\right>=0$.
\item As a corollary, integration by parts can be performed and one can flip derivatives, e.g., $\left<R_\alpha^\mu\nabla_\mu S_\beta\right> = -\left<S_\beta\nabla_\mu R_\alpha^\mu\right>$.
\end{itemize}
Therefore, to obtain the backreaction one has to calculate
\begin{equation}
\label{eq:aveom}
\left<G_{\mu\nu}\right> + \left<\Delta G_{\mu\nu}^{NL}\right> = 8\pi G \left<T_{\mu\nu}\right>
\end{equation}
up to second order in $h$ (higher orders are extremely small)\footnote{When performing the scalar-vector-tensor decomposition to second order in perturbation theory, one has to take into account the contributions from the coupling between scalar and tensor perturbations \cite{Marozzi:2014xma}. These contributions are automatically being taken into account here as we are not decomposing the metric perturbation and everything is given in terms of the entire perturbation $h_{\mu\nu}$.}. Taking the average of Equation \eqref{eq:mod}, gives
\begin{align}
\label{eq:modav}
-\xi\left<\Delta G_{\mu\nu}^\alpha\right> &= 2\left(\left<R_{\mu\nu}\log\left(\frac{\Box}{\mu^2}\right)R\right>-\frac14 \bar g_{\mu\nu}\left<R\log\left(\frac{\Box}{\mu^2}\right)R\right>\right)\nonumber\\
& - 2\left<(\nabla_\mu\nabla_\nu-g_{\mu\nu}\Box)\log\left(\frac{\Box}{\mu^2}\right)R\right>.
\end{align}
It follows from the rules that
\begin{equation}
\left<R_{\mu\nu}\log\left(\frac{\Box}{\mu^2}\right)R^{\mu\nu}\right> = \bar R_{\mu\nu}\log\left(\frac{\Box}{\mu^2}\right)\bar R^{\mu\nu} + \left<R_{\mu\nu}^{(1)}\log\left(\frac{\Box}{\mu^2}\right)R^{(1)\mu\nu}\right>,
\label{eq:avricci}
\end{equation}
since the average of linear terms in $h$ vanishes. Cross terms (e.g. $\bar R R^{(2)}$) are absent as they are negligible due to the condition \eqref{eq:con}. In addition, the last line of Equation \eqref{eq:modav} has a global derivative so that the high-frequency contribution averages to zero.

The combination of Equations \eqref{eq:modav} and \eqref{eq:avricci} results in
\begin{align}
\label{eq:av1}
-\xi\left<\Delta G_{\mu\nu}^\alpha\right> &= 2\left(\bar R_{\mu\nu} -\frac14 \bar g_{\mu\nu}\bar R\right)\log\left(\frac{\Box}{\mu^2}\right)\bar R + 2\left(\left<R_{\mu\nu}^{(1)}\log\left(\frac{\Box}{\mu^2}\right)R^{(1)}\right> - \frac14 \bar g_{\mu\nu} \left<R^{(1)}\log\left(\frac{\Box}{\mu^2}\right)R^{(1)}\right>\right)\nonumber\\
&-2(\nabla_\mu\nabla_\nu-\bar g_{\mu\nu}\Box)\log\left(\frac{\Box}{\mu^2}\right)\bar R.
\end{align}
Similarly, taking the average of Equation \eqref{eq:mod1} gives
\begin{align}
\label{eq:av2}
\zeta\left<\Delta G_{\mu\nu}^\beta\right> &= -\frac12 \bar g_{\mu\nu}\left(\bar R_{\rho\sigma}\log\left(\frac{\Box}{\mu^2}\right)\bar R^{\rho\sigma} + \left<R_{\rho\sigma}^{(1)}\log\left(\frac{\Box}{\mu^2}\right)R^{(1)\rho\sigma}\right>\right)\nonumber\\
& + \Box\log\left(\frac{\Box}{\mu^2}\right)\bar R_{\mu\nu} + \bar g_{\mu\nu}\nabla_\rho\nabla_\sigma\log\left(\frac{\Box}{\mu^2}\right)\bar R^{\rho\sigma}\nonumber\\
& + \bar R_\mu^\sigma\log\left(\frac{\Box}{\mu^2}\right)\bar R_{\nu\sigma} + \bar R_\nu^\sigma\log\left(\frac{\Box}{\mu^2}\right)\bar R_{\mu\sigma} +  2\left<R_\mu^{(1)\sigma}\log\left(\frac{\Box}{\mu^2}\right)R_{\nu\sigma}^{(1)}\right>\nonumber\\
& - \nabla_\rho\nabla_\mu \log\left(\frac{\Box}{\mu^2}\right) \bar R^{\rho}_\nu - \nabla_\rho\nabla_\nu \log\left(\frac{\Box}{\mu^2}\right) \bar R_\mu^{\rho}.
\end{align}

Combining Equations \eqref{eq:aveom}, \eqref{eq:av1} and \eqref{eq:av2} leads to the background EOM
\begin{align}
\label{eq:bkeom}
&\bar R_{\mu\nu}-\frac12 \bar g_{\mu\nu}\bar R -\frac{2}{\xi}\left[\left(\bar R_{\mu\nu}-\frac14 \bar g_{\mu\nu}\bar R\right)\log\left(\frac\Box{\mu^2}\right)\bar R - (\nabla_\mu\nabla_\nu-\bar g_{\mu\nu}\Box)\log\left(\frac{\Box}{\mu^2}\right)\bar R\right]\nonumber\\
& -\frac1{2\zeta} \bar g_{\mu\nu}\bar R_{\rho\sigma}\log\left(\frac{\Box}{\mu^2}\right)\bar R^{\rho\sigma} + \frac1{\zeta}\bar R_\mu^\sigma\log\left(\frac{\Box}{\mu^2}\right)\bar R_{\nu\sigma} + \frac1{\zeta}\bar R_\nu^\sigma\log\left(\frac{\Box}{\mu^2}\right)\bar R_{\mu\sigma} + \frac1{\zeta}\Box\log\left(\frac{\Box}{\mu^2}\right)\bar R_{\mu\nu}\nonumber\\
& + \frac1{\zeta}\bar g_{\mu\nu}\nabla_\rho\nabla_\sigma\log\left(\frac{\Box}{\mu^2}\right)\bar R^{\rho\sigma} - \frac1\zeta\nabla_\rho\nabla_\mu \log\left(\frac{\Box}{\mu^2}\right) \bar R^{\rho}_\nu - \frac1\zeta\nabla_\rho\nabla_\nu \log\left(\frac{\Box}{\mu^2}\right) \bar R_\mu^{\rho}\nonumber\\
&= 8\pi G(\left<T_{\mu\nu}\right> + t_{\mu\nu}^{GR} +  t_{\mu\nu}^{NL}),
\end{align}
where $t_{\mu\nu}^{GR}$ is the classical contribution to the GW stress-energy tensor:
\begin{equation}
t_{\mu\nu}^{GR} = -\frac{1}{8\pi G}\left(\left<R_{\mu\nu}^{(2)}\right>-\frac{1}{2}\bar g_{\mu\nu}\left<R^{(2)}\right>\right)
\end{equation}
and $t_{\mu\nu}^{NL}$ is the non-local one:
\begin{align}
\label{eq:em}
t_{\mu\nu}^{NL} &= -\frac{1}{8\pi G}\Bigg[-\frac2\xi\left(\left<R_{\mu\nu}^{(1)}\log\left(\frac{\Box}{\mu^2}\right)R^{(1)}\right>-\frac{1}{4}\bar g_{\mu\nu}\left<R^{(1)}\log\left(\frac{\Box}{\mu^2}\right)R^{(1)}\right>\right)\nonumber\\
&+\frac2{\zeta}\left<R_\mu^{(1)\sigma}\log\left(\frac{\Box}{\mu^2}\right)R_{\nu\sigma}^{(1)}\right>-\frac1{2\zeta}\bar g_{\mu\nu}\left<R_{\rho\sigma}^{(1)}\log\left(\frac{\Box}{\mu^2}\right)R^{(1)\rho\sigma}\right>\Bigg].
\end{align}
Similarly, the local contribution is given by
\begin{align}
t_{\mu\nu}^L &= -\frac{1}{8\pi G}\Bigg[-32\pi G b_1\left(\left<R_{\mu\nu}^{(1)}R^{(1)}\right>-\frac{1}{4}\bar g_{\mu\nu}\left<R^{(1)}R^{(1)}\right>\right)\nonumber\\
&+32\pi G b_2\left<R_\mu^{(1)\sigma}R_{\nu\sigma}^{(1)}\right>-8\pi G b_2\bar g_{\mu\nu}\left<R_{\rho\sigma}^{(1)}R^{(1)\rho\sigma}\right>\Bigg].
\end{align}
Therefore, the total GW stress-energy tensor is $t_{\mu\nu} = t_{\mu\nu}^{GR} + t_{\mu\nu}^{L} + t_{\mu\nu}^{NL}$.

At this point some comments are necessary. First of all, observe that the left-hand side of Equation \eqref{eq:bkeom} corresponds solely to the background effect, which we interpret as pure gravity. In fact, the left-hand side is exactly the same as in Equation \eqref{eq:eom} when $g_{\mu\nu}$ is replaced by $\bar g_{\mu\nu}$. The right-hand side represents the matter sector, as usual, but with the inclusion of the GW contribution. Such a contribution represents the most general stress-energy tensor to leading order, accounting for both classical and quantum effects. Note that, due to the diffeomorphism invariance of the theory, the total energy-momentum tensor is covariantly conserved
\begin{equation}
\nabla^\mu(T_{\mu\nu}+t_{\mu\nu})=0,
\end{equation}
which shows that energy and momentum are exchanged between matter sources and GWs. Far away from the source, this gives the conservation of the GW energy-momentum tensor
\begin{equation}
\partial^\mu t_{\mu\nu} = 0.
\end{equation}

Up to this point, no gauge conditions have been applied and $t_{\mu\nu}$ also accounts for spurious degrees of freedom. To eliminate them, we shall take the limit where the GW is far away from the source, so that the background is nearly flat and the linear EOM becomes \cite{Calmet:2017rxl}
\begin{equation}
\Box_\eta h_{\mu\nu} + 16\pi G\left[b_2 + \beta \log\left(\frac{\Box_\eta}{\mu^2}\right)\right]\Box_\eta^2 h_{\mu\nu} = 0,
\label{eq:flat}
\end{equation}
where $\Box_\eta=\eta^{\mu\nu}\partial_\mu\partial_\nu$ is the flat D'Alembert operator. Note the absence of the parameter $\alpha$ in Equation \eqref{eq:flat}. This happens because $\alpha$ is proportional to terms depending on the trace $h$, which can be taken as zero far away from the source. Using the gauge conditions $\partial^\nu h_{\mu\nu}=0$ and $h=0$ (only valid outside the source) together with Equation \eqref{eq:flat} in the definition of $t_{\mu\nu}$ gives
\begin{equation}
t_{\mu\nu} = \frac{1}{8\pi G}\left[\frac14 \left<\partial_\mu h_{\alpha\beta}\partial_\nu h^{\alpha\beta}\right> + \frac12\left<h_\mu^\sigma\Box_\eta h_{\nu\sigma}\right> - \frac18 \eta_{\mu\nu} \left<h_{\rho\sigma}\Box_\eta h^{\rho\sigma}\right>\right],
\label{eq:qtgr}
\end{equation}
Comparing this to Equation \eqref{eq:gr}, it is clearly seen that the first term in $t_{\mu\nu}$ corresponds to GR, while the other two come from quantum corrections. Observe that the log operators do not appear explicitly in Equation \eqref{eq:qtgr} as the gravitational field is on shell. This means that their contribution will only show up in the field solutions. For the same reason, the procedure \eqref{eq:log} of imposing causality need not be pursued at this stage as the non-local effects are only reflected in the solutions for $h_{\mu\nu}$. The parameters $b_2$ and $\beta$ now only appear in the mass $m$ of $h_{\mu\nu}$.

\begin{figure}
\centering
\includegraphics[scale=0.4]{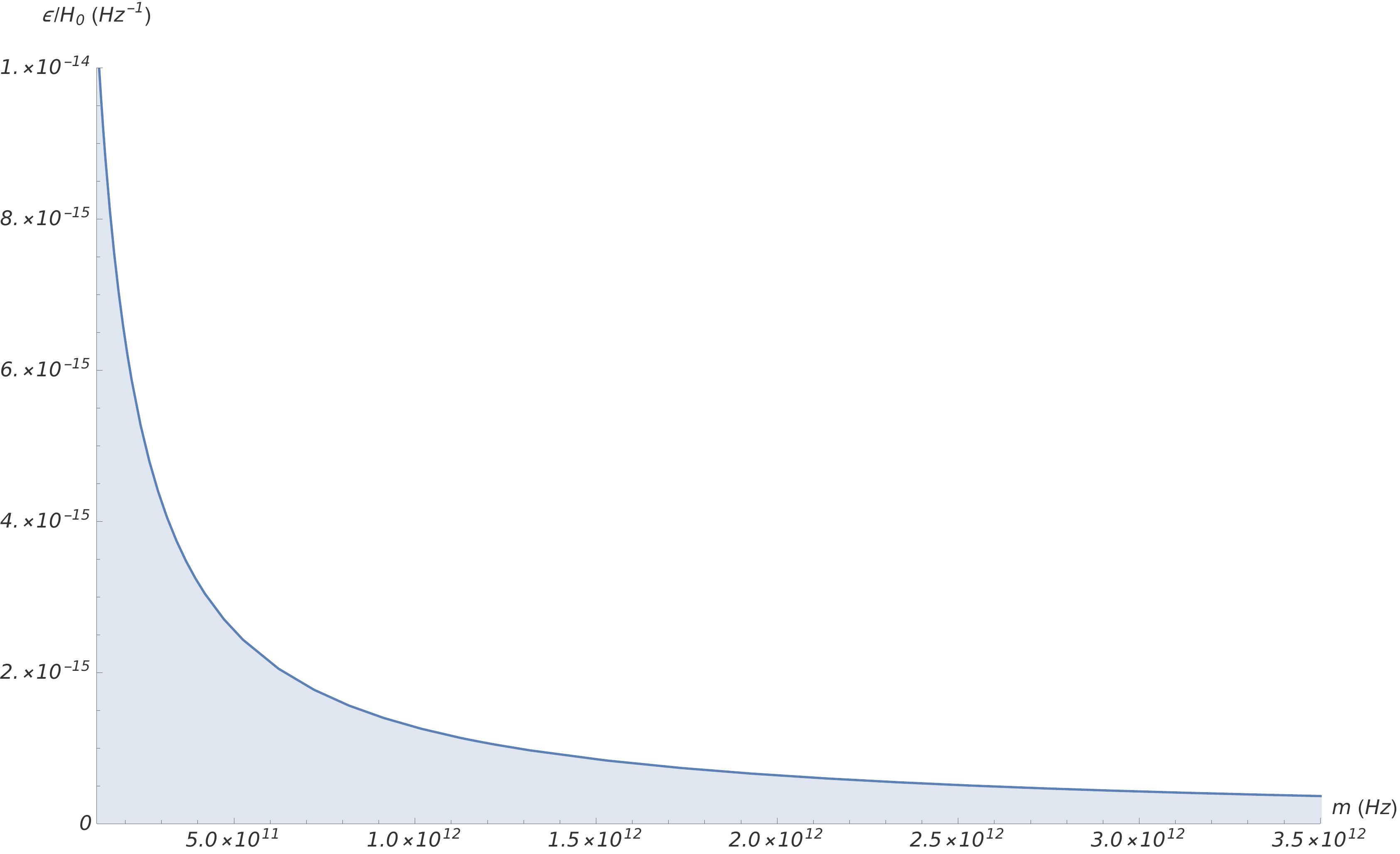}
\caption{The blue area in the graph represents the allowed region of the parameter space $(m,\epsilon)$.}
\label{fig}
\end{figure}

The GW energy density is then
\begin{equation}
\rho\equiv t_{00} = \frac{1}{8\pi G}\left[\frac14\left<\dot h_{\alpha\beta}\dot h^{\alpha\beta}\right> + \frac12\left<h_0^\alpha\Box_\eta h_{0\alpha}\right> + \frac18\left<h_{\rho\sigma}\Box_\eta h^{\rho\sigma}\right>\right].
\label{eq:dens}
\end{equation}
As a concrete example, take a plane wave solution propagating in the $z$ direction
\begin{equation}
h_{\mu\nu} = \epsilon_{\mu\nu}\cos(\omega t - k z).
\label{eq:pw}
\end{equation}
Plugging this into Equation \eqref{eq:dens} gives
\begin{equation}
\rho = \frac{1}{16\pi G}\left[\frac{\epsilon^2\omega^2}{4} + \frac{1}{2}\left(\epsilon_0^\alpha \epsilon_{0\alpha} + \frac{\epsilon^2}{4}\right)(\omega^2-k^2)\right].
\label{eq:densplane}
\end{equation}
Therefore, modifications in the dispersion relations lead to measurable differences into the GW energy. In the case of the classical wave, i.e. $\omega^2=k^2$, the second term vanishes identically, resulting in the classical energy as expected. In the most general case, there could be complex frequencies leading to damping as was shown in \cite{Calmet:2016sba, Calmet:2014gya, Calmet:2016fsr, Calmet:2017omb}. In such case, Equation \eqref{eq:densplane} can be straightforwardly generalized. Note that the second term in \eqref{eq:densplane} is proportional to the particle's mass $m$ and, therefore, is constant as any change in the frequency would be compensated by a change in the momentum. Dividing the constant term by the critical density $\rho_c=\frac{3H_0^2}{8\pi G}$, where $H_0$ is the today's Hubble constant, leads to the frequency-independent gravitational wave density parameter $\Omega_0$ which was constrained to be smaller than $1.7\times 10^{-7}$ by LIGO \cite{TheLIGOScientific:2016dpb}:
\begin{equation}
\Omega_0 = \frac{1}{12}\left(\epsilon_0^\alpha \epsilon_{0\alpha} + \frac{\epsilon^2}{4}\right)\frac{m^2}{H_0^2} < 1.7\times 10^{-7}.
\end{equation}
We remind the reader that the initial parameters $b_2$ and $\beta$ appear only in terms of the mass $m$ as the field $h_{\mu\nu}$ is on shell. Figure \ref{fig} shows the allowed region of the parameter space $(m,\epsilon)$.
Using the lower bound on the mass of the complex pole\footnote{This conservative bound, and consequently the bound on $\epsilon$, was obtained assuming all the energy of a merger goes into the complex mode. Naturally, this does not represent the real situation as the classical mode should also be produced. In a more careful analysis, we expect to get a better bound.} found in \cite{Calmet:2016sba}, i.e. $m>5\times 10^{-13}$GeV, we can translate the above constraint to
\begin{equation}
\epsilon < 1.4\times 10^{-33}.
\end{equation}
To the best of our knowledge, this is the first bound ever found on the amplitude of the massive mode. It is 12 orders of magnitude smaller than the strain sensibility of LIGO's interferometer, which can probe amplitudes up to $\sim 10^{-22}$ in the frequency range from 10 Hz to 10 kHz. Although it seems hopeless to reach such small distances, the Chongqing University detector (currently under development) will be able to probe amplitudes as small as $10^{-32}$ \cite{Baker:2009zzb} in the high-frequency range 0.1--10 GHz, which is not far from the bound just found. Observe, however, that we have found an upper bound on $\epsilon$ and not a lower one, thus $\epsilon$ could be arbitrarily small and not be detectable by the Chongqing detector. Should the existence of these extra modes be confirmed in future experiments, this would be the first evidence for a massive mode.

As it was stressed before, the effective energy-momentum tensor may lead to a contribution to the accelerated expansion of today's universe if its trace is different than zero. The trace of the GW energy-momentum tensor \eqref{eq:qtgr} is non-vanishing:
\begin{equation}
t = -\frac{1}{32\pi G}\left<h_{\alpha\beta}\Box_\eta h^{\alpha\beta}\right>\neq 0,
\end{equation}
as the gravitational field now satisfies the modified EOM \eqref{eq:flat}. Therefore, the energy-momentum tensor $t_{\mu\nu}$ can be split into a traceful and a traceless component
\begin{equation}
t_{\mu\nu} = t_{\mu\nu}-\frac14 \eta_{\mu\nu}t + \frac14 \eta_{\mu\nu}t
\end{equation}
and the cosmological constant can be identified as
\begin{equation}
\Lambda\equiv \frac{1}{16}\left<h_{\alpha\beta}\Box_\eta h^{\alpha\beta}\right> = \frac{1}{16}\epsilon^2 m^2,
\label{eq:cc}
\end{equation}
where in the second equality the plane wave solution \eqref{eq:pw} was used. After taking the average, $\Lambda$ depends very slowly on space and time. In fact, it is precisely constant accross any region of length $d$ and its variation only becomes appreciable in a region containing several lenghts of size $d$. Therefore, for our purposes, we can safely neglect the spacetime dependence of the emergent cosmological constant $\Lambda$ and consider it a constant indeed. Remember that, initially, the cosmological constant was set to zero. A non-zero initial or bare cosmological constant $\Lambda_b$ would just be shifted by the $\Lambda$ found above and the physical cosmological constant would be $\Lambda_{eff}\equiv\Lambda_b + \Lambda$. The important proposition here is that quantum gravitational waves give a non-zero contribution to the cosmological constant $\Lambda_{eff}$. In this scenario, the new gravitational interactions and degrees of freedom appearing in high energies are represented by non-local effects in the low-energy limit. The latter, combined with the local interactions, yields a gravitational energy-momentum tensor whose trace is non-vanishing and which contributes to the total cosmological constant.

\section{Gravitational wave propagation}
\label{sec:prop}
Up to now, only the physics of the low-frequency waves has been considered. For completeness, we shall turn our attention to the high-frequency ones, which will lead to the equation describing the GW propagation in curved spacetime. This is easily achieved by subtracting the background equation \eqref{eq:bkeom} from the total EOM \eqref{eq:eom}
\begin{equation}
G_{\mu\nu} + \Delta G_{\mu\nu} - \left<G_{\mu\nu} + \Delta G_{\mu\nu}\right> = 8\pi G(T_{\mu\nu} - \left<T_{\mu\nu}\right>),
\end{equation}
where $\Delta G_{\mu\nu} = \Delta G_{\mu\nu}^L + \Delta G_{\mu\nu}^{NL}$. Ignoring the local part for a moment and keeping only the terms up to linear order in $h$ and $\lambda/L$ gives
\begin{align}
& R_{\mu\nu}^{(1)} - \frac12 \bar g_{\mu\nu}R^{(1)} + \frac2\xi(\nabla_\mu\nabla_\nu-\bar g_{\mu\nu}\Box)\log\left(\frac\Box{\mu^2}\right)R^{(1)} + \frac1\zeta \Bigg[ \Box\log\left(\frac\Box{\mu^2}\right)R^{(1)}_{\mu\nu}\nonumber\\
& + \bar g_{\mu\nu}\nabla_\rho\nabla_\sigma \log\left(\frac\Box{\mu^2}\right)R^{(1)\rho\sigma} - \nabla_\rho\nabla_\mu\log\left(\frac\Box{\mu^2}\right)R^{(1)\rho}_\nu - \nabla_\rho\nabla_\nu\log\left(\frac\Box{\mu^2}\right)R^{(1)\rho}_\mu\Bigg]=0
\label{eq:prop}
\end{align}
Outside the matter source, we can use the gauge $\nabla^\nu h_{\mu\nu}=0$ together with $h=0$, leading to
\begin{equation}
\Box h_{\mu\nu} + 16\pi G\beta\log\left(\frac{\Box}{\mu^2}\right)\Box^2 h_{\mu\nu} = 0.
\end{equation}
Analogously, including the local terms gives
\begin{equation}
\Box h_{\mu\nu} + 16\pi G\left[b_2 + \beta\log\left(\frac{\Box}{\mu^2}\right)\right]\Box^2 h_{\mu\nu} = 0.
\label{eq:propc}
\end{equation}
When deriving Equation \eqref{eq:propc}, we made use of the commutation relation of covariant derivatives and we discarted terms proportional to the background curvature as they only contribute to higher orders in $\lambda/L$. Equation \eqref{eq:propc} describes the propagation of GWs in an arbitrary curved background in the absence of external matter, when the only source for a non-vanishing Ricci tensor is the GW energy-momentum tensor. The curvature terms do not appear as they provide no contribution to leading order. Therefore, the case where curvature is present can be obtained by applying a simple ``minimal coupling'' prescription to Equation \eqref{eq:flat} where spacetime is flat, that is, by performing the following substitution
\begin{align}
\eta_{\mu\nu}\to \bar g_{\mu\nu},\\
\partial_\mu\to \nabla_\mu.
\end{align}
Equations \eqref{eq:bkeom} and \eqref{eq:propc} describe together the entire classical and quantum process (to leading order) of the GW self-gravitation: small perturbations around spacetime change the curvature, which in turn modify the GW's trajectory and vice-versa.

\section{Conclusions}
\label{sec:conc}
We showed in this paper how to calculate the quantum corrections to the GW stress-energy tensor. The result shows that quantum effects promote the traceless tensor $t_{\mu\nu}^{\text{GR}}$ to a traceful quantity that contributes to the current accelerated expansion of the universe. In addition, the energy density component acquires a dependence on modifications to the dispersion relation, indicating a useful observable to probe when looking for quantum gravitational effects. In fact, by using the latest LIGO's data, it was obtained a new upper bound on the amplitude of the massive mode. We also showed that the high-frequency mode equation led to a generalization of the wave equation \eqref{eq:flat} to arbitrary curved spacetimes \eqref{eq:propc}. Such generalization is important to the study of quantum GW solutions in cosmology and in the early universe where the spacetime was highly curved. Lastly, it must be stressed once again that these quantum contributions are model independent (since they are derived from an Effective Field Theory) and constitute actual predictions of Quantum Gravity, shedding new light on Quantum Gravity as a whole and giving us some hints of how a complete theory, if such theory exists, should behave below the Planck scale.

\noindent{\it Acknowledgments:}
This work is supported by the National Council for Scientific and Technological Development (CNPq - Brazil).

%%%%%%%%%%%%%%%%%%%%%%%%%%%%%%%%%%%%%%%%%%%%%%%%%%%%%%%%%%%%%%%%%
%%%
%%%                     BIBLIOGRAPHY
%%%
%%%%%%%%%%%%%%%%%%%%%%%%%%%%%%%%%%%%%%%%%%%%%%%%%%%%%%%%%%%%%%%%%

\bigskip{}


\baselineskip=1.6pt

\begin{thebibliography}{10}

%\cite{Abbott:2016blz}
\bibitem{Abbott:2016blz} 
  B.~P.~Abbott {\it et al.} [LIGO Scientific and Virgo Collaborations],
  %``Observation of Gravitational Waves from a Binary Black Hole Merger,''
  Phys.\ Rev.\ Lett.\  {\bf 116}, no. 6, 061102 (2016)
  doi:10.1103/PhysRevLett.116.061102
  [arXiv:1602.03837 [gr-qc]].
  %%CITATION = doi:10.1103/PhysRevLett.116.061102;%%
  %1272 citations counted in INSPIRE as of 07 Apr 2017

%\cite{Stein:2010pn}
\bibitem{Stein:2010pn} 
  L.~C.~Stein and N.~Yunes,
  %``Effective Gravitational Wave Stress-energy Tensor in Alternative Theories of Gravity,''
  Phys.\ Rev.\ D {\bf 83}, 064038 (2011)
  doi:10.1103/PhysRevD.83.064038
  [arXiv:1012.3144 [gr-qc]].
  %%CITATION = doi:10.1103/PhysRevD.83.064038;%%
  %29 citations counted in INSPIRE as of 16 Mar 2017

%\cite{Preston:2016sip}
\bibitem{Preston:2016sip} 
  A.~W.~H.~Preston,
  %``Cosmological backreaction in higher-derivative gravity expansions,''
  JCAP {\bf 1608}, no. 08, 038 (2016)
  doi:10.1088/1475-7516/2016/08/038
  [arXiv:1605.06121 [gr-qc]].
  %%CITATION = doi:10.1088/1475-7516/2016/08/038;%%
  %1 citations counted in INSPIRE as of 16 Mar 2017

%\cite{Preston:2014tua}
\bibitem{Preston:2014tua} 
  A.~W.~H.~Preston and T.~R.~Morris,
  %``Cosmological back-reaction in modified gravity and its implications for dark energy,''
  JCAP {\bf 1409}, 017 (2014)
  doi:10.1088/1475-7516/2014/09/017
  [arXiv:1406.5398 [gr-qc]].
  %%CITATION = doi:10.1088/1475-7516/2014/09/017;%%
  %5 citations counted in INSPIRE as of 16 Mar 2017

%\cite{Saito:2012xa}
\bibitem{Saito:2012xa} 
  K.~Saito and A.~Ishibashi,
  %``High frequency limit for gravitational perturbations of cosmological models in modified gravity theories,''
  PTEP {\bf 2013}, 013E04 (2013)
  doi:10.1093/ptep/pts061
  [arXiv:1209.5159 [gr-qc]].
  %%CITATION = doi:10.1093/ptep/pts061;%%
  %4 citations counted in INSPIRE as of 16 Mar 2017

%\cite{Berry:2011pb}
\bibitem{Berry:2011pb} 
  C.~P.~L.~Berry and J.~R.~Gair,
  %``Linearized f(R) Gravity: Gravitational Radiation and Solar System Tests,''
  Phys.\ Rev.\ D {\bf 83}, 104022 (2011)
  Erratum: [Phys.\ Rev.\ D {\bf 85}, 089906 (2012)]
  doi:10.1103/PhysRevD.85.089906, 10.1103/PhysRevD.83.104022
  [arXiv:1104.0819 [gr-qc]].
  %%CITATION = doi:10.1103/PhysRevD.85.089906, 10.1103/PhysRevD.83.104022;%%
  %62 citations counted in INSPIRE as of 03 Apr 2017

%\cite{Rasanen:2010fe}
\bibitem{Rasanen:2010fe}
  S.~Rasanen,
  %``Backreaction as an alternative to dark energy and modified gravity,''
  arXiv:1012.0784 [astro-ph.CO].
  %%CITATION = ARXIV:1012.0784;%%
  %10 citations counted in INSPIRE as of 03 Apr 2017

%\cite{Rasanen:2003fy}
\bibitem{Rasanen:2003fy} 
  S.~Rasanen,
  %``Dark energy from backreaction,''
  JCAP {\bf 0402}, 003 (2004)
  doi:10.1088/1475-7516/2004/02/003
  [astro-ph/0311257].
  %%CITATION = doi:10.1088/1475-7516/2004/02/003;%%
  %168 citations counted in INSPIRE as of 03 Apr 2017

%\cite{Buchert:2011sx}
\bibitem{Buchert:2011sx} 
  T.~Buchert and S.~Rasanen,
  %``Backreaction in late-time cosmology,''
  Ann.\ Rev.\ Nucl.\ Part.\ Sci.\  {\bf 62}, 57 (2012)
  doi:10.1146/annurev.nucl.012809.104435
  [arXiv:1112.5335 [astro-ph.CO]].
  %%CITATION = doi:10.1146/annurev.nucl.012809.104435;%%
  %99 citations counted in INSPIRE as of 03 Apr 2017

%\cite{Donoghue:2014yha}
\bibitem{Donoghue:2014yha} 
  J.~F.~Donoghue and B.~K.~El-Menoufi,
  %``Nonlocal quantum effects in cosmology: Quantum memory, nonlocal FLRW equations, and singularity avoidance,''
  Phys.\ Rev.\ D {\bf 89}, no. 10, 104062 (2014)
  doi:10.1103/PhysRevD.89.104062
  [arXiv:1402.3252 [gr-qc]].
  %%CITATION = doi:10.1103/PhysRevD.89.104062;%%
  %28 citations counted in INSPIRE as of 16 Mar 2017

%\cite{Calmet:2017rxl}
\bibitem{Calmet:2017rxl} 
  X.~Calmet, S.~Capozziello and D.~Pryer,
  %``Gravitational Effective Action at Second Order in Curvature and Gravitational Waves,''
  arXiv:1708.08253 [hep-th].
  %%CITATION = ARXIV:1708.08253;%%

%\cite{Codello:2015pga}
\bibitem{Codello:2015pga} 
  A.~Codello and R.~K.~Jain,
  %``On the covariant formalism of the effective field theory of gravity and its cosmological implications,''
  Class.\ Quant.\ Grav.\  {\bf 34}, no. 3, 035015 (2017)
  doi:10.1088/1361-6382/aa549d
  [arXiv:1507.07829 [astro-ph.CO]].
  %%CITATION = doi:10.1088/1361-6382/aa549d;%%
  %10 citations counted in INSPIRE as of 16 Mar 2017

%\cite{Donoghue:2015nba}
\bibitem{Donoghue:2015nba} 
  J.~F.~Donoghue and B.~K.~El-Menoufi,
  %``Covariant non-local action for massless QED and the curvature expansion,''
  JHEP {\bf 1510}, 044 (2015)
  doi:10.1007/JHEP10(2015)044
  [arXiv:1507.06321 [hep-th]].
  %%CITATION = doi:10.1007/JHEP10(2015)044;%%
  %12 citations counted in INSPIRE as of 15 Aug 2017

%\cite{Isaacson:1968zza}
\bibitem{Isaacson:1968zza} 
  R.~A.~Isaacson,
  %``Gravitational Radiation in the Limit of High Frequency. II. Nonlinear Terms and the Ef fective Stress Tensor,''
  Phys.\ Rev.\  {\bf 166}, 1272 (1968).
  doi:10.1103/PhysRev.166.1272
  %%CITATION = doi:10.1103/PhysRev.166.1272;%%
  %250 citations counted in INSPIRE as of 16 Mar 2017

%\cite{Isaacson:1967zz}
\bibitem{Isaacson:1967zz} 
  R.~A.~Isaacson,
  %``Gravitational Radiation in the Limit of High Frequency. I. The Linear Approximation and Geometrical Optics,''
  Phys.\ Rev.\  {\bf 166}, 1263 (1967).
  doi:10.1103/PhysRev.166.1263
  %%CITATION = doi:10.1103/PhysRev.166.1263;%%
  %250 citations counted in INSPIRE as of 16 Mar 2017

%\cite{Marozzi:2014xma}
\bibitem{Marozzi:2014xma} 
  G.~Marozzi and G.~P.~Vacca,
  %``Tensor Mode Backreaction During Slow-roll Inflation,''
  Phys.\ Rev.\ D {\bf 90}, no. 4, 043532 (2014)
  doi:10.1103/PhysRevD.90.043532
  [arXiv:1405.3933 [gr-qc]].
  %%CITATION = doi:10.1103/PhysRevD.90.043532;%%
  %4 citations counted in INSPIRE as of 09 Dec 2017

%\cite{Calmet:2016sba}
\bibitem{Calmet:2016sba} 
  X.~Calmet, I.~Kuntz and S.~Mohapatra,
  %``Gravitational Waves in Effective Quantum Gravity,''
  Eur.\ Phys.\ J.\ C {\bf 76}, no. 8, 425 (2016)
  doi:10.1140/epjc/s10052-016-4265-8
  [arXiv:1607.02773 [hep-th]].
  %%CITATION = doi:10.1140/epjc/s10052-016-4265-8;%%
  %2 citations counted in INSPIRE as of 16 Mar 2017

%\cite{Calmet:2014gya}
\bibitem{Calmet:2014gya} 
  X.~Calmet,
  %``The Lightest of Black Holes,''
  Mod.\ Phys.\ Lett.\ A {\bf 29}, no. 38, 1450204 (2014)
  doi:10.1142/S0217732314502046
  [arXiv:1410.2807 [hep-th]].
  %%CITATION = doi:10.1142/S0217732314502046;%%
  %15 citations counted in INSPIRE as of 11 Aug 2017

%\cite{Calmet:2016fsr}
\bibitem{Calmet:2016fsr} 
  X.~Calmet and I.~Kuntz,
  %``Higgs Starobinsky Inflation,''
  Eur.\ Phys.\ J.\ C {\bf 76}, no. 5, 289 (2016)
  doi:10.1140/epjc/s10052-016-4136-3
  [arXiv:1605.02236 [hep-th]].
  %%CITATION = doi:10.1140/epjc/s10052-016-4136-3;%%
  %10 citations counted in INSPIRE as of 11 Aug 2017

%\cite{Calmet:2017omb}
\bibitem{Calmet:2017omb} 
  X.~Calmet, R.~Casadio, A.~Y.~Kamenshchik and O.~V.~Teryaev,
  %``Graviton propagator, renormalization scale and black-hole like states,''
  arXiv:1708.01485 [hep-th].
  %%CITATION = ARXIV:1708.01485;%%

%\cite{TheLIGOScientific:2016dpb}
\bibitem{TheLIGOScientific:2016dpb} 
  B.~P.~Abbott {\it et al.} [LIGO Scientific and Virgo Collaborations],
  %``Upper Limits on the Stochastic Gravitational-Wave Background from Advanced LIGO’s First Observing Run,''
  Phys.\ Rev.\ Lett.\  {\bf 118}, no. 12, 121101 (2017)
  doi:10.1103/PhysRevLett.118.121101
  [arXiv:1612.02029 [gr-qc]].
  %%CITATION = doi:10.1103/PhysRevLett.118.121101;%%
  %8 citations counted in INSPIRE as of 24 May 2017

%\cite{Baker:2009zzb}
\bibitem{Baker:2009zzb} 
  R.~M.~L.~Baker,
  %``The Peoples Republic of China High-Frequency Gravitational Wave research program,''
  AIP Conf.\ Proc.\  {\bf 1103}, 548 (2009).
  doi:10.1063/1.3115564
  %%CITATION = doi:10.1063/1.3115564;%%

\end{thebibliography}
\end{document}